\documentclass[aps,prl,noshowpacs,superscriptaddress,amsmath,amssymb,twocolumn,preprintnumbers]{revtex4}
\usepackage{xspace}
\usepackage{graphicx}
\usepackage{dcolumn}
\usepackage{epsf}
\usepackage{bm}
\usepackage[english]{babel}
\usepackage{bbold}
\usepackage{graphics}
\usepackage{hyperref}
\usepackage{xurl}
\usepackage{pstricks}
\usepackage{color}
\usepackage{relsize,exscale,scalefnt,anyfontsize,cases}
\usepackage[utf8]{inputenc}
\usepackage{dsfont}

\def\be{\begin{equation}}
\def\ee{\end{equation}}
\def\bea{\begin{eqnarray}}
\def\eea{\end{eqnarray}}
 
\newcommand{\Nc}{\ensuremath{N_c}\xspace}
\newcommand{\eq}[1]{(\ref{#1})}
\newcommand{\ave}[1]{\langle{#1}\rangle}
\newcommand{\R}{\textnormal{R}}
\newcommand{\Kvec}{{\boldsymbol K}}
\newcommand{\pt}{p_{_\perp}}
\newcommand{\dd}{{\rm d}}
\newcommand{\Phat}{\hat{{\cal P}}}
\newcommand{\ella}{L_{_{\rm A}}}
\newcommand{\pp}{\ensuremath{\text{pp}}\xspace}
\newcommand{\pA}{\ensuremath{\text{pA}}\xspace}

\begin{document}
\title{Quenching of light hadron spectra in pA collisions from fully coherent energy loss}
\author{Fran\c{c}ois Arleo}
\affiliation{Laboratoire Leprince-Ringuet, \'Ecole polytechnique, Institut polytechnique de Paris, CNRS/IN2P3, 91128 Palaiseau, France}
\author{St\'ephane Peign\'e}
\affiliation{SUBATECH UMR 6457 (IMT Atlantique, Universit\'e de Nantes, IN2P3/CNRS), 4 rue Alfred Kastler, 44307 Nantes, France}

\date{\today}

\begin{abstract}
We single out the role of fully coherent induced gluon radiation on light hadron production in pA collisions. The effect has an interesting color structure, as the induced radiation depends on the global color charge of the partonic subprocess final state. Baseline predictions for light hadron nuclear suppression in pPb collisions at the LHC are provided, taking into account only the effect of fully coherent energy loss, which proves of the same order of magnitude as gluon shadowing or saturation. This underlines the need to include fully coherent energy loss in phenomenological studies of hadron production in pA collisions.
\end{abstract}

\maketitle

\setcounter{footnote}{0}
\renewcommand{\thefootnote}{\arabic{footnote}}

The wealth of hadron production data in proton-nucleus collisions at collider (RHIC, LHC) energies allows for a detailed study of parton dynamics in cold nuclear matter. Over the last two decades, several approaches, in particular the Color Glass Condensate~\cite{Gelis:2010nm} or the use of nuclear parton distribution functions (nPDFs) within collinear factorization~\cite{Armesto:2006ph}, have been used in phenomenological studies of hadron production in high-energy pA collisions~\cite{Tribedy:2011aa,Albacete:2012xq,Rezaeian:2012ye,Lappi:2013zma,Helenius:2012wd,QuirogaArias:2010wh,Kusina:2017gkz,Eskola:2019bgf}. To our knowledge, none of these approaches have yet addressed the role of fully coherent energy loss (FCEL) in cold nuclear matter discussed throughout this Letter. FCEL is expected in all processes where the underlying partonic process, when viewed in the target nucleus rest frame, consists in forward scattering of an incoming energetic parton to an outgoing color charge~\cite{Arleo:2010rb,Peigne:2014uha} or a colorful system of partons~\cite{Peigne:2014rka}. It arises from induced gluon radiation with formation time $t_{\textnormal{f}}$ much larger than the medium length, $t_{\textnormal{f}} \gg L$, and associated to an energy spectrum $\omega \, {\dd I}/{\dd\omega}$ scaling in $x \equiv \omega/E$, with $E$ and $\omega$ the energies of the incoming parton and of the induced radiation, respectively. As a result, the average parton energy loss scales in $E$, $\Delta E_{_{\textnormal{FCEL}}} \propto E$~\cite{Arleo:2010rb}, thus overwhelming the LPM energy loss $\Delta E_{_{\textnormal{LPM}}} \propto L^2$~\cite{Baier:1996sk,Baier:1996kr,Zakharov:1996fv,Zakharov:1997uu}. FCEL is predicted from first principles in various formalisms~\cite{Arleo:2010rb,Armesto:2012qa,Arleo:2012rs,Armesto:2013fca,Peigne:2014uha,Peigne:2014rka,Liou:2014rha,Munier:2016oih}, including the saturation formalism~\cite{Liou:2014rha,Munier:2016oih}, and has been shown to be a key effect in quarkonium nuclear suppression in pA collisions~\cite{Arleo:2012hn,Arleo:2012rs,Arleo:2013zua}. It is thus natural to investigate the effect of FCEL in other processes, such as open heavy flavour, light hadron, or jet production in pA collisions. In the present study we focus on single inclusive light hadron production.

Our primary goal is to set a baseline for the quantitative role of FCEL in light hadron nuclear suppression, by taking into account {\it only} this effect. We show that similarly to quarkonium production, the effect of FCEL on light hadron production in pA collisions is sizable, yet with novel features that will be underlined. In particular, FCEL depends on the global color charge of the parton pair produced in the partonic subprocess. As an interesting consequence, single hadron pA production is sensitive to the color states ($\textnormal{SU}(\Nc)$ irreducible representations) of the parton pair, and thus to unusual color factors. Another goal of our study is to minimize the model dependence of FCEL baseline predictions. To that end, the pp cross section is not taken from theory but determined from a fit to the data, and we thus predict the hadron nuclear modification factor $R_{\pA}$ (due to FCEL), rather than absolute cross sections. As a result, our predictions of a significant hadron suppression arising from FCEL also have a remarkably small associated uncertainty. 
 
These observations indicate that FCEL should be taken into account in phenomenological interpretations of the pA data. It has been suggested to use present and future data on hadron production in pA collisions ($h^\pm$~\cite{Helenius:2012wd,QuirogaArias:2010wh}, $D$/$B$ mesons~\cite{Kusina:2017gkz,Eskola:2019bgf}, quarkonia~\cite{Kusina:2017gkz}) as a reliable probe of nPDFs (and of saturation~\cite{Albacete:2012xq}), assuming other physical effects to be negligible. Our study shows that the latter assumption should be reconsidered, due to the presence of sizable FCEL effects. In particular, FCEL should be included in nPDF global fit analyses that use hadron production pA data. In pA collisions, electroweak processes where FCEL is absent~\cite{Arleo:2010rb} should be preferred for a direct extraction of nPDFs, as for instance weak boson~\cite{Paukkunen:2010qg} and Drell-Yan~\cite{Arleo:2015qiv} production. Hadron production at the future EIC~\cite{Accardi:2012qut,EIC} would also be a direct probe of nPDFs and saturation since no FCEL is expected in deep inelastic scattering~\cite{Arleo:2010rb}. 
  
In order to implement FCEL in quarkonium hadroproduction~\cite{Arleo:2012rs,Arleo:2013zua}, the differential production cross section in pA is obtained from that in pp collisions by {\it rescaling} the quarkonium energy by a factor $1+x$, where $x$ denotes the fractional energy loss, which is equivalent to a rapidity shift $\delta \equiv \ln{(1+x)}$,  
\be
\label{rapidity-shift-jpsi}
\frac{1}{A}\frac{\dd\sigma_{\pA}^{\psi}(y)}{\dd y} = \int_0^{\delta_{\rm max}}  \! \! \dd{\delta} \, \, \Phat (x) \, \frac{\dd\sigma_{\pp}^{\psi}\left(y+\delta \right)}{\dd y} \, .
\ee
In Eq.~\eq{rapidity-shift-jpsi}, $\delta_{\textnormal{max}}=\min(\ln{2}, y_{\textnormal{max}}-y)$, with $y_{\textnormal{max}}$ the maximal quarkonium rapidity, and the \emph{quenching weight} $\hat{\cal P}(x = e^{\delta}-1)$ is the fractional energy loss probability distribution associated to the production of a color octet compact $Q \bar{Q}$ pair~\cite{Arleo:2012rs,Arleo:2013zua}. The quenching weight is related to the induced coherent spectrum $\dd I/\dd x$ associated to $g \to g$ forward scattering (with a final `massive gluon') derived and discussed in Refs.~\cite{Arleo:2010rb,Arleo:2012rs,Peigne:2014uha}, 
\be
\label{quenching-spectrum}
\hat{\cal P} (x) = \frac{\partial}{\partial x} \, \exp \left\{ - \int_{x}^{\infty} \dd x^\prime \left. {\frac{\dd{I}}{\dd x^\prime}}\right|_{g\to g} \right\} \, .
\ee 
This approach has been successful in describing the quarkonium suppression reported in pA collisions from SPS to LHC energies~\cite{Arleo:2012rs,Arleo:2013zua}. 

\begin{figure}[t]
\centering
\includegraphics[width=8cm]{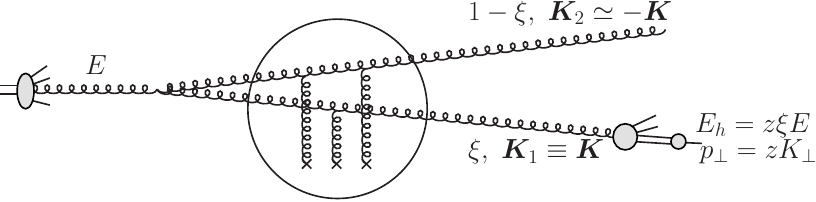}
\caption{Contribution to light hadron production in pA collisions from the partonic process $g + {\rm A} \to gg + {\rm X}$ followed by gluon fragmentation $g \to h$, as viewed in the target nucleus rest frame.}
\label{fig:typical-process}
\end{figure}

The procedure to implement FCEL in light hadron production in pA collisions is analogous to quarkonium production, Eq.~\eq{rapidity-shift-jpsi}, despite new features. We will consider single inclusive hadron production at `large' $\pt \gg \ell_{_{\perp \rm A}}$, where $\ell_{_{\perp \rm A}}$ is the transverse momentum broadening of a fast gluon across the target A. It is related to the average path length $\ella$ as $\ell_{_{\perp \rm A}}^2 = \hat{q} \ella$, with $\hat{q}$ the transport coefficient in cold nuclear matter, parametrized as $\hat{q} \equiv \hat{q}_{_0} \left( {10^{-2}}/{x_{_2}} \right)^{0.3}$~\cite{ACP,Arleo:2012rs}. 
When viewed in the target rest frame, hadron production at $\pt \gg \ell_{_{\perp \rm A}}$ arises from $1 \to 2$ forward scattering (at leading-order), where the partons of the final pair (`dijet' in the following) have transverse momenta $\Kvec_1 \equiv \Kvec$ and $\Kvec_2 \simeq -\Kvec$, with $K_{_\perp} \equiv |\Kvec| \gg \ell_{_{\perp \rm A}}$, and energy fractions $\xi$ and $1-\xi$ with respect to the incoming parton energy $E$. This $1 \to 2$ scattering is followed by the fragmentation of one parton of the pair into the tagged hadron, which thus inherits the transverse momentum $\pt = z K_{_\perp}$, where $z$ is the fragmentation variable.
The process is illustrated in Fig.~\ref{fig:typical-process} in the case of $g \to gg$ scattering. Although the present framework is generalized in Ref.~\cite{ACP} to other processes ($q\to qg$ and $g\to q\bar{q}$), in this Letter we focus on $g \to gg$ which dominates light hadron production around mid-rapidity at the LHC~\cite{Sassot:2010bh}.
The $g\to q\bar{q}$ channel will be generalized in a future study to the production of 
{\it massive} quarks, $g \to Q\bar{Q}$, allowing for the calculation of FCEL effects on open heavy flavour production, recently measured at forward rapidity by LHCb~\cite{Aaij:2019lkm,LHCb:2019dpz}.

In general, calculating explicitly the induced coherent radiation spectrum associated to the production of a dijet (or multi-parton system) may be complicated. When the induced radiation does not probe the color charges of the dijet constituents, however, the FCEL spectrum depends only on the dijet global color charge. 
Within this `pointlike dijet approximation'~\footnote{We have verified that the condition for this approximation to hold, $\ln{(\ell_{_{\perp \rm A}}^2/x^2 K_{_\perp}^2)} \gg 1$~\cite{Peigne:2014rka,ACP}, is fulfilled in our calculation.}, the induced spectrum off a dijet in color state $\R$  must coincide with that associated to $1 \to 1$ scattering~\cite{Arleo:2010rb,Arleo:2012rs,Peigne:2014uha} applied to the case of an outgoing particle of Casimir $C_{_{\R}}$ and mass equal to the dijet mass $M_{\xi} = K_{_\perp}/\sqrt{\xi(1-\xi)}$, namely, 
\be
\label{PDAspec-1}
\left. x \frac{\dd I_{_\R}}{\dd x}\right|_{g\to (gg)_{_\R}} = \, C_{_{\R}} \, \frac{\alpha_s}{\pi} \,\ln{\left(\frac{1+ \ell_{_{\perp \rm A}}^2 / x^2 M_{\xi}^2}{1+ \ell_{_{\perp \rm p}}^2 / x^2 M_{\xi}^2 } \right)} \, .
\ee 

In the pointlike dijet approximation, implementing FCEL is simply achieved by separating the different dijet color states in the hadron production cross section, and performing the rapidity shift for each dijet color state separately~\cite{ACP}. Note that for a pointlike dijet, the same rapidity shift applies to the dijet, its constituent partons, and the tagged hadron.
Two ingredients are needed: (i) the probabilities $\rho_{_{\R}}$ for the dijet (digluon) to be in color state $\R$. Those are determined in Ref.~\cite{ACP} from the $g \to gg$ scattering amplitude written as a linear combination of projectors on the various color states of a gluon pair~\cite{Dokshitzer:2005ig}, and turn out to depend only on $\xi$; (ii) the FCEL quenching weight $\hat{\cal P}_{_\R}$ corresponding to a pointlike dijet in color state $\R$, obtained by replacing in Eq.~\eq{quenching-spectrum} the quantity $\left.{\dd I}/{\dd x}\right|_{g\to g}$ by $\left.{\dd I_{_\R}}/{\dd x}\right|_{g\to (gg)_{_\R}}$ given in \eq{PDAspec-1}. 

The hadron production cross section in pA collisions is thus written as~\cite{ACP}
\bea
\frac{1}{A} \, \frac{\dd \sigma_{\rm pA}^{h}(y,\pt)}{\dd y \, \dd \pt} = \sum_{\R}  \, \int_0^{\delta_{\rm max}} \! \! \dd{\delta} \, \int_0^1 \dd{\xi} \hskip 18mm &&  \nonumber \\ 
\times \,  \rho_{_{\R}}(\xi) \, \Phat_{_\R}(x, M_{\xi}) \,  \frac{\dd\sigma_{\pp}^{h}(y+\delta, \pt, \xi)}{\dd y \, \dd \pt \dd{\xi}} \, , && 
\label{sig-pA-y} 
\eea
extending \eq{rapidity-shift-jpsi} to the case of light hadron production. In Eq.~\eq{sig-pA-y}, the quenching weight is evaluated at $x= e^{\delta}-1$ and its dependence on $M_{\xi}$ is made explicit, as well as the dependence of the pp and pA cross sections on the hadron $\pt$. Let us emphasize that our implementation of FCEL effects consists in a different organization of the perturbative expansion as compared to next-to-leading order (NLO) approaches evaluating absolute pA cross sections~\cite{Chirilli:2011km,Kang:2014lha,Ducloue:2016shw,Fujii:2013gxa,Ducloue:2015gfa,Ma:2015sia}. The latter studies in principle account for the induced radiation of a single gluon, as part of all NLO corrections. In our approach, the induced radiation is resummed to all orders through the expression of the quenching weight. 

Using \eq{sig-pA-y}, we obtain the nuclear production ratio in minimum bias pA collisions as compared to pp collisions, 
\bea
\label{RpA-y} 
R_{\pA}^{h}(y,\pt)    
= \sum_{\R} \, \int_0^{\delta_{\textnormal{max}}} \! \!  \dd{\delta} \, \, {\ave{\rho_{_{\R}}(\xi) \,  \Phat_{_\R}(x, M_{\xi})}}_{y+\delta, \, \pt} &&  \nonumber \\
\times \  \frac{\dd \sigma_{\rm pp}^{h}(y+\delta,\pt)}{\dd y\, \dd \pt} \, \Big/ \,  \frac{\dd \sigma_{\rm pp}^{h}(y,\pt)}{\dd y\, \dd \pt} \, ,  \hskip 5mm  && 
\eea
where  $\ave{ \ }_{_{y, \pt}}$ denotes the $\xi$-average in pp dijet events where a hadron of rapidity $y$ and transverse momentum $\pt$ is produced~\cite{ACP}. 

In the spirit of the FCEL studies on quarkonium production~\cite{Arleo:2012rs,Arleo:2013zua,Arleo:2014oha}, the goal is to determine hadron suppression with a minimal model-dependence. We use the following procedure, which introduces a limited set of parameters, the latter being then varied for a proper determination of theoretical uncertainties. 

Defining the parameter $\bar{\xi}$ as the value which fulfils ${\ave{\rho_{_{\R}}(\xi)\, \Phat_{_{\R}}(x, M_{\xi})}}_{y+\delta,\,\pt}=\rho_{_{\R}}(\bar{\xi})\,\Phat_{_{\R}}(x,M_{\bar{\xi}})$ (which existence is guaranteed by the mean value theorem), the nuclear production ratio~\eq{RpA-y} can be written as
\bea
\label{RpA-y-master}
R_{\pA}^{h}(y, \pt, \bar{\xi}) = \sum_{\R} \,  \rho_{_{\R}}(\bar{\xi})  \, R_{\pA}^{\R}(y, \pt, \bar{\xi})  \, ,   \hskip 10mm && \\ 
\label{RpA-yR}
R_{\pA}^{\R}(y, \pt,\bar{\xi}) = \int_0^{\delta_{\rm max}} \! \! \dd{\delta} \, \, \Phat_{_{\R}}(x, M_{\bar{\xi}}) \, \frac{\ \frac{\dd \sigma_{\rm pp}^{h}(y+\delta,\,\pt)}{\dd y\, \dd \pt}\ }{\ \frac{\dd \sigma_{\rm pp}^{h}(y,\,\pt)}{\dd y\, \dd \pt}\ } \, \, . \hskip 0mm &&
\eea 
Thus, $R_{\pA}^{h}$ is the color average of the nuclear modification factors $R_{\pA}^{\R}$ corresponding to a hadron produced from a dijet in color state $\R$. The light hadron nuclear suppression due to FCEL is computed using Eqs.~\eq{RpA-y-master}-\eq{RpA-yR}. The uncertainty associated to the choice of $\bar{\xi}$ will be estimated by taking $\bar{\xi}=0.50\pm0.25$, the default value $\bar{\xi}=0.50$ corresponding to the symmetric configuration of two jets of equal rapidity (which is most likely according to dihadron correlation measurements~\cite{Khachatryan:2016txc}), and the $\bar{\xi}$ variation range corresponding to a rapidity difference between the two jets of approximately $\Delta y=\pm \ln(\bar{\xi}/(1-\bar{\xi}))\simeq \pm1$ unit. 

The pp cross section entering Eq.~\eq{RpA-yR} is not taken from theory but from a fit to the data using a simple analytic form, $\dd\sigma_{\pp}^{h}/\dd\pt\!\dd y = f(\pt) \times \left(1- \frac{2\,\pt}{\sqrt{s}} \cosh{y} \right)^{n}$~\cite{Arleo:2013zua}, where only the rapidity dependent factor, and thus the parameter $n$, is relevant when computing $R_{\pA}^{h}$ from Eqs.~\eq{RpA-y-master}-\eq{RpA-yR}. Using CMS double differential measurements in pPb collisions at $\sqrt{s}=5.02$~TeV leads to the value $n=15\pm5$~\cite{ACP}. 
The remaining parameters are the transport coefficient normalization $\hat{q}_{_0}$, and the fragmentation variable $z$ (determining $K_{_\perp} = \pt/z$), both entering only the expression of the spectrum \eq{PDAspec-1}. We vary $\hat{q}_{_0}$ in the range $0.07$--$0.09$~GeV$^2$/fm (with a default value $0.075$~GeV$^2$/fm)~\cite{Arleo:2012rs},  and take $z = 0.7 \pm 0.2$ based on NLO calculations of hadron production at the LHC~\cite{Sassot:2010bh}. Finally, we set $\alpha_s=0.5$, $L_{_\textnormal{Pb}}=10.11$~fm, and $L_{_\textnormal{p}}=1.5$~fm~\cite{Arleo:2012rs}.
In order to estimate the theoretical uncertainties, the quantities $n$, $\bar{\xi}$, $z$, and $\hat{q}_{_0}$ are varied around their default values, which define the central prediction. Assuming the parameters to be uncorrelated, the uncertainty band of the model predictions is determined using the Hessian method~\cite{Pumplin:2001ct} (applied in the context of FCEL effects on quarkonium production in~\cite{Arleo:2014oha}) from the variation of each parameter around its default value while keeping the other parameters fixed.

\begin{figure}[t]
\centering
\includegraphics[width=7.5cm]{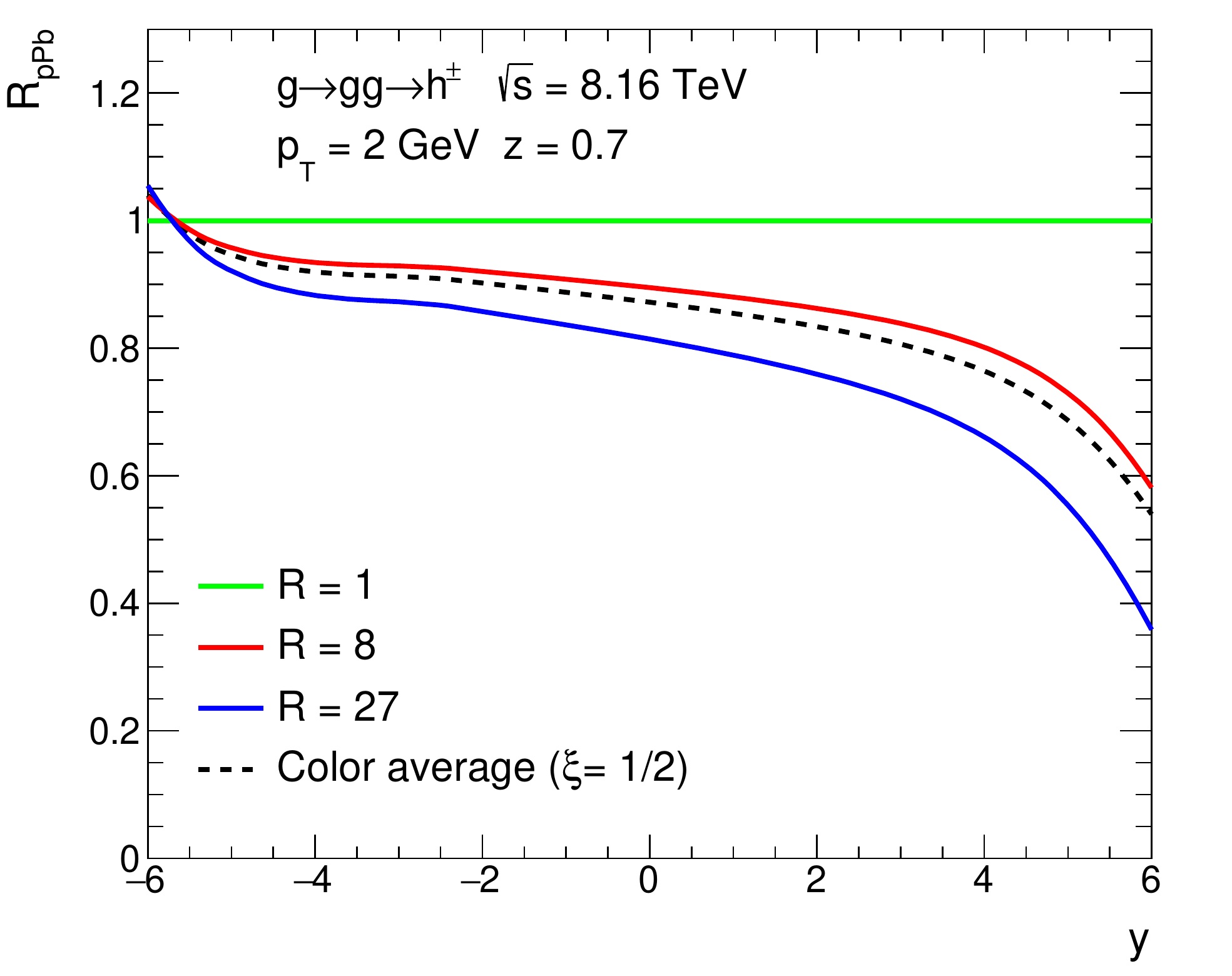}
\caption{Rapidity dependence of  $R_{\pA}^{\R}$, Eq.~\eq{RpA-yR}, at $\pt = 2$~GeV, in the $g\to (gg)_{_\R}$ channel for $\R={\bf 1}$ (green line), $\R={\bf 8}$ (red line), and $\R={\bf 27}$ (blue line). The color-averaged nuclear modification factor $R_{\pA}^{h}$, Eq.~\eq{RpA-y-master}, is shown for $\xi=1/2$ (dashed black line).}
\label{fig-RpA-gg}
\end{figure}

Calculations of light hadron suppression in pPb collisions at current top LHC energy, $\sqrt{s}=8.16$~TeV, are shown here as a function of $\pt$ and $y$. Although somewhat academic, it is instructive to discuss first the nuclear production ratio \eq{RpA-yR} for a final gluon pair in a given color state $\R$, and then obtain the `inclusive' hadron suppression through the average \eq{RpA-y-master} over color states. In Fig.~\ref{fig-RpA-gg} we show the rapidity dependence of $R_{\pA}^{\R}$ at fixed $\pt=2$~GeV, for the color states $R={\bf 1}, {\bf 8}, {\bf 27}$ (having non-zero probabilities~\cite{ACP}), as well as the color average (dashed line). When the $gg$ final state is color singlet, the induced gluon spectrum \eq{PDAspec-1} vanishes, and no FCEL effect is expected in this channel, $R_{\pA}^{{\bf 1}}=1$. More interesting is the suppression of the octet $gg$ final state, which shape is reminiscent of the suppression predicted for quarkonium~\cite{Arleo:2012rs}. The suppression gets quite strong at large $y$, $R_{\pA}^{\bf 8}\simeq 0.6$ at $y=6$, due to the steeply falling pp cross section. Note that the steeply {\it rising} pp cross section at very backward rapidity leads to a slight enhancement below $y\simeq-6$. For the 27-plet $gg$ final state, the suppression expected from FCEL follows the same pattern, but is more pronounced due to the larger Casimir, $C_{27}=2\,(\Nc+1)$, in the prefactor of the induced gluon spectrum \eq{PDAspec-1}.

\begin{figure}[t]
\centering
\includegraphics[width=7.5cm]{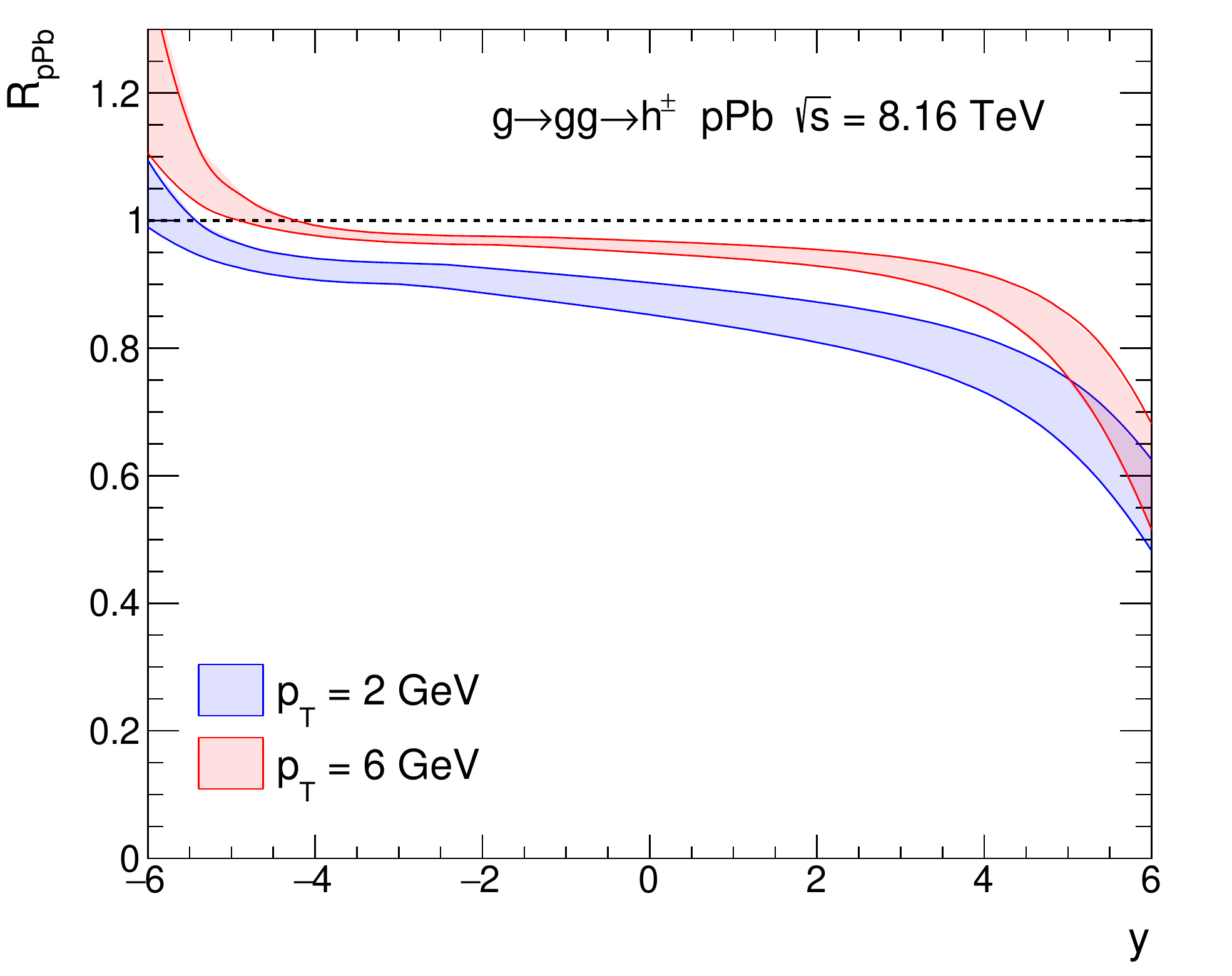}  \\
\includegraphics[width=7.5cm]{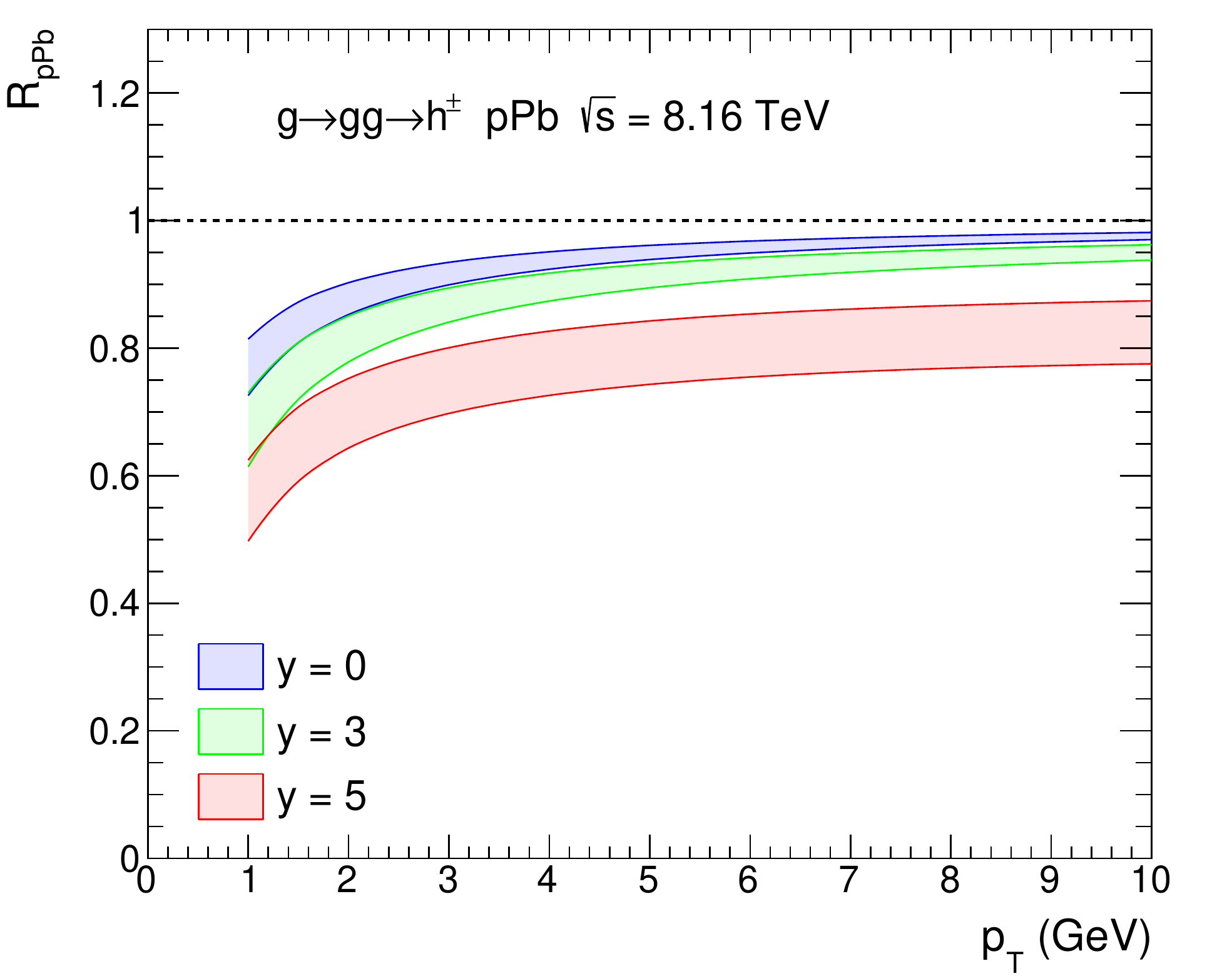}  
\caption{FCEL effects on $R_{\pA}^{h}$ as a function of $y$ (top) and $\pt$ (bottom), for different values of $\pt$ and $y$ respectively, in pPb collisions at $\sqrt{s}=8.16$~TeV.}
\label{fig-RPA-y-pt}
\end{figure}

The light hadron suppression obtained after averaging over color states is shown with its uncertainty band in Fig.~\ref{fig-RPA-y-pt} (top) as a function of $y$, for $\pt=2$~GeV and $\pt=6$~GeV. Because of the specific dependence of the induced gluon spectrum \eqref{PDAspec-1} in $M_{_\xi} \propto K_{_\perp} = \pt / z$, FCEL effects weaken at larger $\pt$, except at very large rapidity, $y \gtrsim 5$, where the effects of the slope of the cross section are larger at higher $\pt$ due to the more restricted phase space. This effect is also visible on the $\pt$ dependence of $R_{\pA}^{h}$ shown in Fig.~\ref{fig-RPA-y-pt} (bottom).  At $y=5$, the suppression flattens for $\pt > 5$~GeV, the effect of the scale dependence being compensated by the strong phase space restriction when $y$ and $\pt$ are both large. 

We stress that the smallness of FCEL relative uncertainties (only $\sim 4\%$ at mid-rapidity) is somehow expected within our approach: FCEL is a medium-induced effect (depending on the difference between coherent radiation spectra in pA and pp collisions), and is thus fully determined within perturbative QCD. Moreover, the parameters $\bar{\xi}$, $z$, and $\hat{q}_{_0}$ enter the induced gluon spectrum \eq{PDAspec-1} through a (large) logarithm, and their variation thus leads to small relative deviations around the central predictions. As for the parameter $n$, its variation affects negligibly the predictions, except at very large $|y|$ where it dominates the total relative uncertainty, which however remains moderate (see Fig.~\ref{fig-RPA-y-pt})~\cite{ACP}. 

\begin{figure}[t]
\centering
\includegraphics[width=7.5cm]{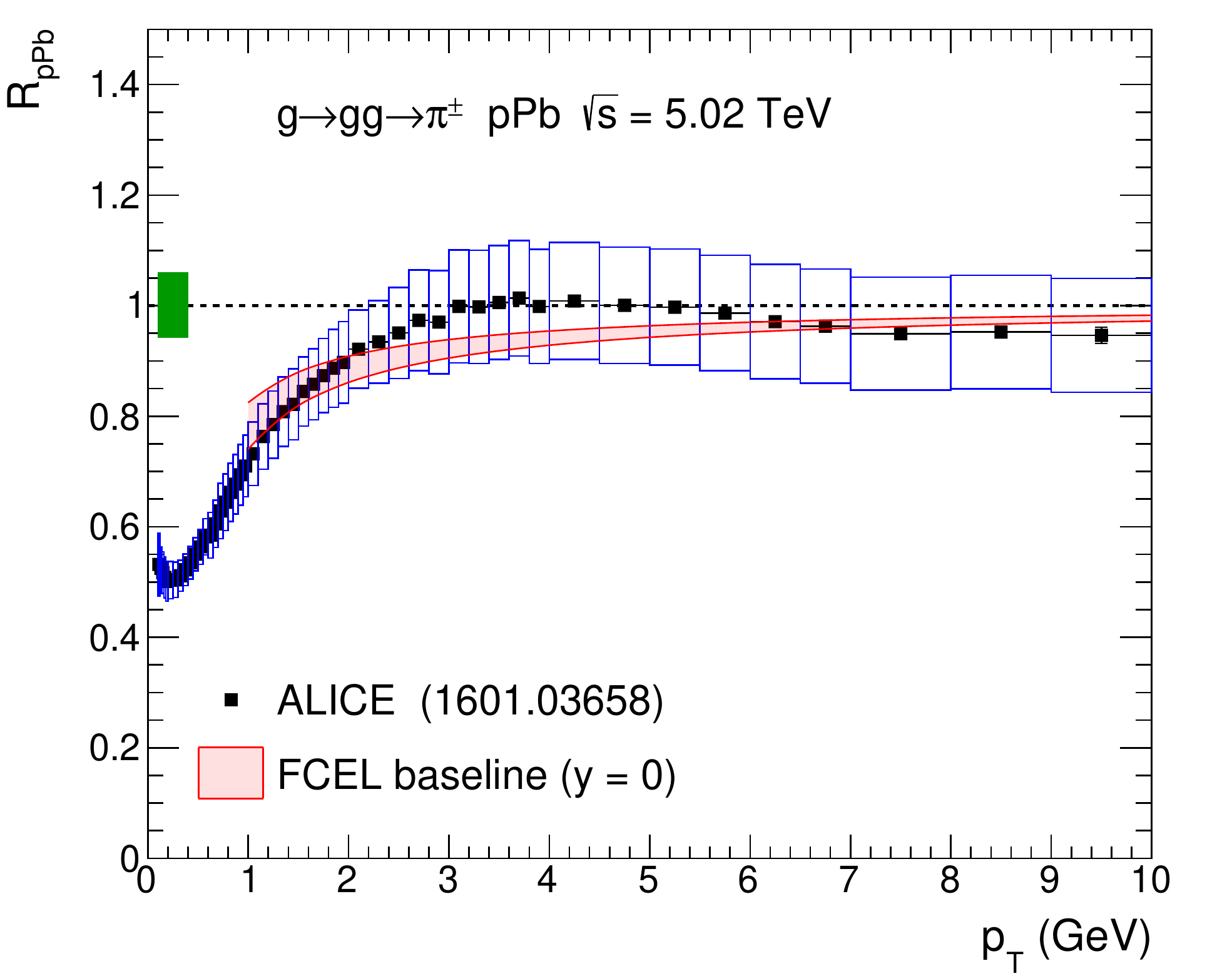} 
\caption{FCEL effects on $R_{\pA}^{h}$ as a function of $y$ in pPb collisions at $\sqrt{s}=5.02$~TeV (red band) in comparison to ALICE charged pion data~\cite{Adam:2016dau}.}
\label{fig-RPA-data}
\end{figure}

Let us now compare the FCEL expectations to the measurements of charged pion suppression in pPb collisions at $\sqrt{s}=5.02$~TeV by the ALICE experiment~\cite{Adam:2016dau}. As shown in Fig.~\ref{fig-RPA-data}, there is a good agreement between data and the FCEL baseline predictions. In particular, the data sometimes attributed to saturation~\cite{Tribedy:2011aa,Albacete:2012xq,Rezaeian:2012ye,Lappi:2013zma}, nPDF effects~\cite{Helenius:2012wd}, and Cronin effect with initial-state energy loss~\cite{Kang:2012kc}, are as well consistent with fully coherent energy loss alone~\footnote{Note however that the significant proton/antiproton enhancement measured by ALICE~\cite{Adam:2016dau}, reminiscent of the Cronin effect, cannot be explained by the sole FCEL effects~\cite{ACP}.}. Since the uncertainties of the FCEL baseline prediction are significantly smaller than those of the measurements, taking into account FCEL should therefore provide strict constraints on other physical processes. It will be interesting to compare FCEL baseline predictions at large rapidity, where FCEL effects become the strongest (see Fig.~\ref{fig-RPA-y-pt}), with future measurements by the LHCb experiment in the rapidity region $2 < y < 4.5$~\cite{Aaij:2014pza}.

In this Letter, FCEL effects on light hadron production in pPb collisions at LHC energies are computed for the first time, and compared to ALICE data at mid-rapidity.  The FCEL effects prove to be of the same order of magnitude as nPDF~\cite{Helenius:2012wd} or saturation~\cite{Tribedy:2011aa,Albacete:2012xq,Rezaeian:2012ye,Lappi:2013zma} effects at mid-rapidity, and moreover subject to quite small uncertainties. This underlines the need to take into account FCEL in phenomenological studies of hadron production in pA collisions. 
This study should pave the way for the systematic computation of FCEL effects in a perturbative QCD calculation of hadron production in pA collisions, through an extension to other partonic subprocesses. The FCEL predictions in the leading-order $g \to q \bar{q}$ and $q \to qg$ channels~\cite{ACP} are qualitatively similar to those in the $g \to gg$ channel discussed here. This program can be achieved at NLO using the FCEL spectrum associated to $1\to 3$ forward scattering~\cite{Peigne:2014rka}, eventually allowing for nPDF NLO global fit analyses taking FCEL effects into account.

\acknowledgments
This work is funded by ``Agence Nationale de la Recherche'', grant ANR-COLDLOSS (ANR-18-CE31-0024-02).

\end{document}